# Super-resolution microscopy in Neurosciences: zoom in on synapses

Laurent COGNET & Brahim LOUNIS,
Univ. Bordeaux, Institut d'Optique & CNRS, Talence, France

**Status**

The remarkably efficient functioning of the brain largely mirrors its multiscale complexity. In particular, it became clear since the emergence of modern neuroscience that basic investigations of brain organization at the subcellular scales are key for the understanding of brain processes. Since then, our knowledge of the cellular mechanisms involved in neuronal communication and their evolution during life has literally exploded. Most of our comprehension comes from the discovery of molecules, genes and signaling cascades that play central roles in the maintenance and plasticity of the neuronal communication. Notably, the identification of the synapse has attracted much attention. It contains key molecules required to induce plastic changes and mediates a large fraction of fast neuronal communication and long-term adaptations. The dimensions of brain synapses are however small, at the submicron scale and therefore they cannot be resolved with conventional optical microscopes. Electron microscopy has been providing ultra-structural information of synapse architectures and to some extent, knowledge about the content and local molecular densities in these structures. However, this knowledge is out-of-reach in living cells and in intact tissues with this imaging modality. In this context, light microscopy constitutes the imaging modality of choice to study synapses in live cell. Continuous developments and refinements of optical imaging techniques delivered an impressive amount of information about synaptic molecule organizations at the nanoscale. A first decisive achievement came from the possibility to detect single molecules in living cells. It not only allows localizing molecules with nanometer accuracies, far below the optical resolution, but also provides the ability to probe the intimate variety of molecule dynamic environments from nano- to micro-scales levels. In the early 2000's the presence of mobile receptors for neurotransmitters was revealed for the first time[2,3] in the postsynaptic membrane of dissociated neuronal cultures and the role of neurotransmitter receptor diffusion in fast synaptic transmission was demonstrated[4]. The advent of several super-resolution methods that followed these early achievements raised great expectations in neurosciences by providing optical images of neuronal structures with unprecedented resolutions.

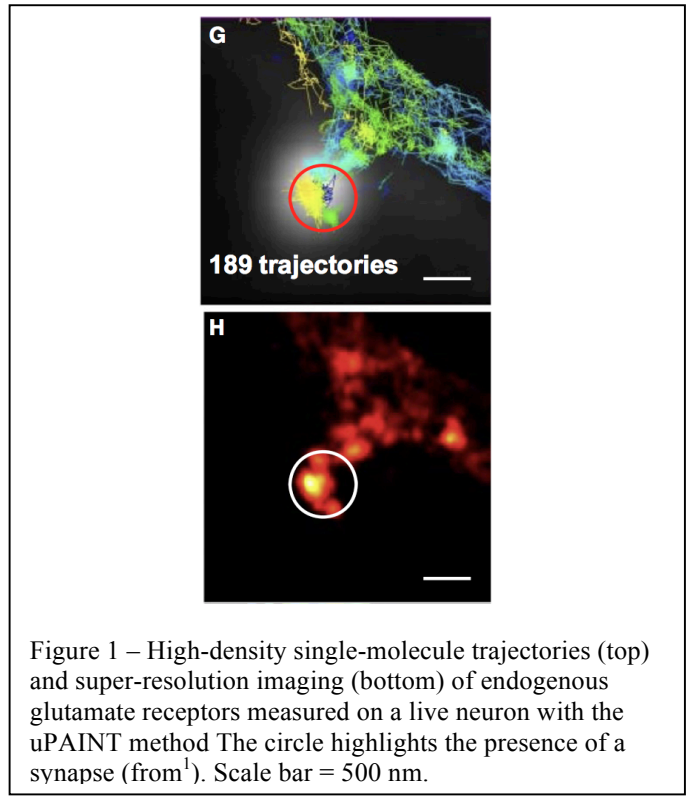

Figure 1 – High-density single-molecule trajectories (top) and super-resolution imaging (bottom) of endogenous glutamate receptors measured on a live neuron with the uPAINT method The circle highlights the presence of a synapse (from[1]). Scale bar = 500 nm.

**Current and Future Challenges**

Super-resolution microscopy methods mostly rely on the control of the number of emitting molecules in specific imaging volumes. Two major types of techniques found major applications in neurosciences: those based on STED where highly localized fluorescence emission volumes are optically shaped and those based on single-molecule localizations ((f)PALM, (d)STORM or (u)PAINT)[5]. Single-molecule localization based methods have proven to be powerful to study nanoscale molecular organizations such as that of postsynaptic receptors and scaffold proteins (e.g., [1,6-8]) but also that of actin and actin binding proteins in axons[9]. On the other hand, STED microscopy allows visualizing dendritic spine shapes in living neurons with remarkable resolution and revealed that spine neck plasticity regulates compartmentalization of synapses[10] and further deciphered the dynamic organization of actin.

As exemplified in the aforementioned achievements, imaging synaptic structures and their molecular contents with nanometer resolutions constitute already a breakthrough for the understanding of synaptic functioning. Because these demonstrations have predominantly been performed on dissociated neuronal cultures or thin fixed brain slices, one of the main challenge will be to transfer such cutting edge technologies to more integrated and ultimately intact living sample. This will allow major fields of neurosciences, such as development, aging and neurodegenerative disorders, to benefit from the unprecedented degree of details provided by these microscopies.

## Advances in Science and Technology to Meet Challenges

The possibility to perform super-resolution imaging in intact living brain tissues to address major neurosciences questions is conditioned to several requirements. Because both light scattering and absorption limit the penetration depth in thick biological samples in the visible domain, super-resolution imaging in tissues is only emerging. Scattering is particularly detrimental for STED microscopy, since the realisation of a zero intensity region for depletion is mandatory. Two-photon microscopy, which uses near-infrared laser excitations, is a common strategy to excite molecules deep in brain tissues. However, due to the high intensities required, photobleaching rates are drastically enhanced and single molecule based super-resolution with two-photon excitation remains difficult. To fully exploit the biological transparency window, near-infrared probes with favourable photophysics still need to be developed, as current red-shifted dyes are not strong emitters.

Another key development will consist in achieving fast wide field super-resolved imaging at rates compatible with the inherent movements of living samples in 3D. Current super-resolution methods are indeed restricted to rather small imaging areas (sub-millimetric), which is constitutive to the novel opportunity to obtain images at superior resolutions. Indeed, by increasing the image resolution by typically a factor 10 to 50 per dimension, the size of a 3D image will be increased by a factor $10^3$ up to $\sim 10^5$. This imposes constrains on the imaging speed as well as data handling. A promising strategy would consist in performing multi-scale imaging where only "relevant" brain sub-areas are imaged at the nanometer scales while the other regions are imaged at lower resolutions. The definition of imaging "relevance" within the brain might depend on time and/or the physio-pathological local state of the sample as well as the nature of the information provided by the low resolution imaging modality. In this context, correlative imaging with other modalities, which do not have to be limited to optical techniques, are interesting routes. Combining multiple super-resolution approaches can also provide brand novel opportunities. Finally, a full understanding of the brain function cannot be obtained solely by images. Manipulation of the brain physio-pathological state will be needed using methodologies that are compatible with super-resolution imaging. Optogenetic methods provide promising tools in this context.

## Concluding Remarks

For widespread application of super-resolution imaging in neuroscience, the main challenge will be to link subcellular information gathered at the molecular scale (e.g. about synaptic processes) to the global organ function obtained at a macroscopic scales, in a well-defined functional state. To this aim, a multidisciplinary effort will be needed where physicist, chemists, computer scientist, neurophysiologists and neuropathologists will have to work together. The task is vertiginous, but is at the level of the complexity of the brain.

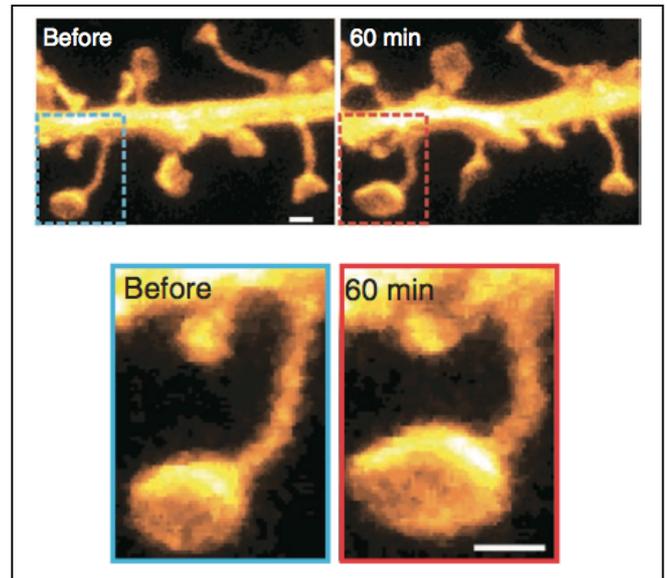

Figure 2 – Effect of long term potentiation on the morphology of synaptic spines and heads revealed by STED microscopy (bottom panels are zoom of top ones. (from[10]). Scale bar = 500 nm.

## References

[1] G. Giannone, E. Hosy, F. Levet, A. Constals, K. Schulze, A. I. Sobolevsky, M. P. Rosconi, E. Gouaux, R. Tampé, D. Choquet, and L. Cognet, Biophys. J. **99,** 1303 (2010).
[2] C. Tardin, L. Cognet, C. Bats, B. Lounis, and D. Choquet, EMBO J. **22,** 4656 (2003).
[3] M. Dahan, S. Levi, C. Luccardini, P. Rostaing, B. Riveau, and A. Triller, Science **302,** 442 (2003).
[4] M. Heine, L. Groc, R. Frischknecht, J.-C. Beique, B. Lounis, G. Rumbaugh, R. L. Huganir, L. Cognet, and D. Choquet, Science **320,** 201 (2008).
[5] A. G. Godin, B. Lounis, and L. Cognet, Biophys J **107,** 1777 (2014).
[6] A. Dani, B. Huang, J. Bergan, C. Dulac, and X. Zhuang, Neuron **68,** 843 (2010).
[7] H. D. MacGillavry, Y. Song, S. Raghavachari, and T. A. Blanpied, Neuron **78,** 615 (2013).
[8] D. Nair, E. Hosy, J. D. Petersen, A. Constals, G. Giannone, D. Choquet, and J. B. Sibarita, J Neurosci **33,** 13204 (2013).
[9] K. Xu, G. Zhong, and X. Zhuang, Science **339,** 452 (2013).
[10] J. Tonnesen, G. Katona, B. Rozsa, and U. V. Nagerl, Nat Neurosci **17,** 678 (2014).